\documentclass[twocolumn, prb]{revtex4}
\pdfoutput=1
\usepackage{psfrag}
\usepackage{graphicx}
\usepackage{amsmath}
\usepackage{amssymb}

\begin{document}
\title{Towards quantitative accuracy in first-principles transport
  calculations: The GW method applied to alkane/gold junctions}
\author{M. Strange}
\affiliation{Center for Atomic-scale Materials Design, Department of
Physics \\ Technical University of Denmark, DK - 2800 Kgs. Lyngby, Denmark}
\author{K. S. Thygesen}
\affiliation{Center for Atomic-scale Materials Design, Department of
Physics \\ Technical University of Denmark, DK - 2800 Kgs. Lyngby, Denmark}
\date{\today}

\begin{abstract}
  The calculation of electronic conductance of nano-scale junctions
  from first principles is a long standing problem in molecular
  electronics. Here we demonstrate excellent agreement with
  experiments for the transport properties of the gold/alkanediamine
  benchmark system when electron-electron interactions are described
  using the many-body GW approximation. The main difference from
  standard density functional theory (DFT) calculations is a
  significant reduction of the contact conductance, $G_c$, due an
  improved alignment of the molecular energy levels with the metal
  Fermi energy. The molecular orbitals involved in the tunneling
  process comprise states delocalized over the carbon backbone and
  states localized on the amine end groups. We find that dynamical
  screening effects renormalize the two types of states in
  qualitatively different ways when the molecule is inserted in the
  junction. Consequently, the GW transport results cannot be mimicked
  by DFT calculations employing a simple scissors operator.
\end{abstract}
\maketitle

\begin{section}{Introduction}
  The conductance of a molecule sandwiched between metallic electrodes
  is sensitive to the chemical and electronic structure of the
  molecule as well as the detailed atomic structure of the metal-molecule
  contact. Variations in the contact geometry beyond experimental control lead
  to an undesired spread in measured conductance properties. For the
  most commonly used thiol anchoring group, these effects are rather
  pronounced due to the many possible contact geometries
  resulting from the large strength of the Au-S interaction.
  Amine groups have been shown to produce more well defined transport
  properties \cite{venkataraman} which can be understood from the
  relatively weak Au-NH$_2$ bond leading to larger structural
  selectivity \cite{iben}.

  Even for a given junction geometry, a quantitatively accurate
  description of electron transport from first principles remains a
  formidable task. Numerous studies based on density functional theory
  (DFT) have shown a significant overestimation of conductance
  relative to experimental values \cite{tomfohr_prb_2002, sen_acsnano_2010, kaun_nanolett_2003, muller_prb_2006, magnus, wohlthat_cpl_2008, mcdermott_jpcc_2009, sheng_jcp_2009,evers,hybertsen,quek,mowbray} (an exception to this
  trend occurs for small molecules, like
  H$_2$ \cite{smit02,thygesen_H2} and
  CO \cite{untiedt04,strange_CO}, which chemisorp strongly to the
  electrodes resulting in resonant transport through broad, partially
  filled resonances). The inability of DFT to describe off-resonant
  tunneling in the simplest molecular junctions limits the predictive
  power of the DFT-based approach to qualitative trends. It is now
  broadly accepted that the failure of DFT is mainly due to its wrong
  description of the molecular energy levels. Indeed, physically
  motivated correction schemes have shown that much improved agreement
  with experiments can be obtained after shifting the DFT molecular
  energy levels \cite{quek,mowbray}. Such corrections are supposed to
  remove the self-interaction errors inherent in standard DFT
  exchange-correlation (xc) functions and account for image charge
  effects induced by the metal contacts. The drawback of the approach
  is that is assumes a weak coupling between molecular orbitals and
  metal states and treats the image plane position as a free
  parameter.

  The (self-consistent) GW approximation \cite{hedin}, which is rooted
  in many-body perturbation theory, was recently found to yield a
  considerable improvement over DFT for the conductance of
  gold/benzenediamine junctions \cite{strange}. Physically, the GW approximation corresponds to Hartree-Fock theory with the bare Coulomb interaction $v=1/|\bf r-\bf r'|$ replaced by a dynamically screened Coulomb interaction $W(\omega)=\epsilon^{-1}(\omega)v$. In contrast to standard
  DFT, the GW approximation is self-interaction free and includes image charge effects due to the metal contacts through the correlation part of the self-energy\cite{neaton,juanma,thygesen_image}. As a consequence, it
  provides quantitatively accurate predictions of energy gaps in
  systems with highly diverse screening properties ranging from
  isolated molecules \cite{rostgaard,blase} over
  semiconductors \cite{hybertsen2}, to metals \cite{barth}. This property
  becomes particularly important for a metal-molecule interface where
  the electronic structure changes from insulating to metallic over a
  few Angstrom. 

  In this work we use the GW approximation to study the role of
  exchange-correlation effects for the energy level alignment and
  electron transport in short alkane chains coupled to gold electrodes
  via amine linker groups. The gold/alkane junction is a benchmark system for molecular charge transport and have been exhaustively investigated experimentally \cite{xu_science_2003, engelkes_jacs_2004, chen_jacs_2006, venkataraman, park_jacs_2007, hybertsen, huang_jacs_2007, widawsky_nanotechnology_2009, kamenetska_prl_2009, song_apl_2007, kim_nano_2011, ruitenbeek, nichols_pccp_2010, haiss_pccp_2009}. We focus here on the amine-linked alkanes to
  avoid the uncertainties related to the gold-thiol contact geometry
  which is presently under debate \cite{ref18,ulstrup,voznyy,ref19,ref20,ref21}. We note that very
  recently it was shown that alkanes can be bound directly to gold
  electrodes without the use of anchoring
  groups \cite{cheng_nature_nanotech_2011}.

  The transport mechanism in (short) saturated molecular wires is
  coherent tunneling via molecular orbitals with energy far from the
  Fermi energy. The conductance versus chain length ($N$) thus follows
  an exponential law of the form
\begin{equation}
G = G_c \exp({-\beta N})
\end{equation}
Recent experimentally reported values for the decay constant $\beta$ of
alkanediamine/gold junctions are in the range 0.9-1.0 per C
atom \cite{venkataraman, hybertsen, park_jacs_2007}, 
but earlier measurements also showed values around
0.8 \cite{chen_jacs_2006}.  Although previous studies based on DFT have yielded
$\beta$ values within the experimental range, the contact conductance, $G_c$, is typically
overestimated by around an order of magnitude \cite{tomfohr_prb_2002, sen_acsnano_2010, kaun_nanolett_2003, muller_prb_2006, magnus, wohlthat_cpl_2008, mcdermott_jpcc_2009, sheng_jcp_2009}.  By comparing DFT
and GW calculations for $N$-alkanediamine molecules with $N=2, 4,6$ we show that the wrong $G_c$ values are a result of incorrect level
alignment in the DFT calculations. Indeed, GW yields a $G_c$ in close
agreement with experimental values. We find a pronounced orbital and
length dependence of the quasiparticle (QP) corrections to the DFT
energies resulting from the different shape and localization of the
molecular orbitals. The QP corrections range from $-0.5$ to $-2.5$~eV and
can be qualitatively explained from a classical image charge model.
\end{section}

\begin{section}{Method}
  The junction geometries were optimized using the real space
  projector augmented wave method GPAW \cite{ref29,GPAW-LCAO} with a grid spacing
  of 0.2~\AA~and the PBE functional for exchange and correlation
  (xc) \cite{ref25}. The molecules were attached to Au(111) surfaces,
  modeled by an eight layer thick $4\times 4$ slab via small four atom
  tips as shown in Figure \ref{fig:figure1}. The surface Brillouin zone
  was sampled on a $4\times 4$ Monkhorst pack $k$-point grid, and the
  structures including molecule, Au tips, and outermost Au surface
  layers were relaxed until the residual force was below 0.03~eV/\AA.
  We have considered $N$-alkanediamine junctions with $N=2$, 4 and 6.
  Key structural parameters can be found here \cite{note}. For
  calculations of the molecules in the gas-phase we include 16~\AA~of
  vacuum between molecules in repeated super cells.
\begin{figure}[!h]
\includegraphics[width=0.95\linewidth, angle=0]{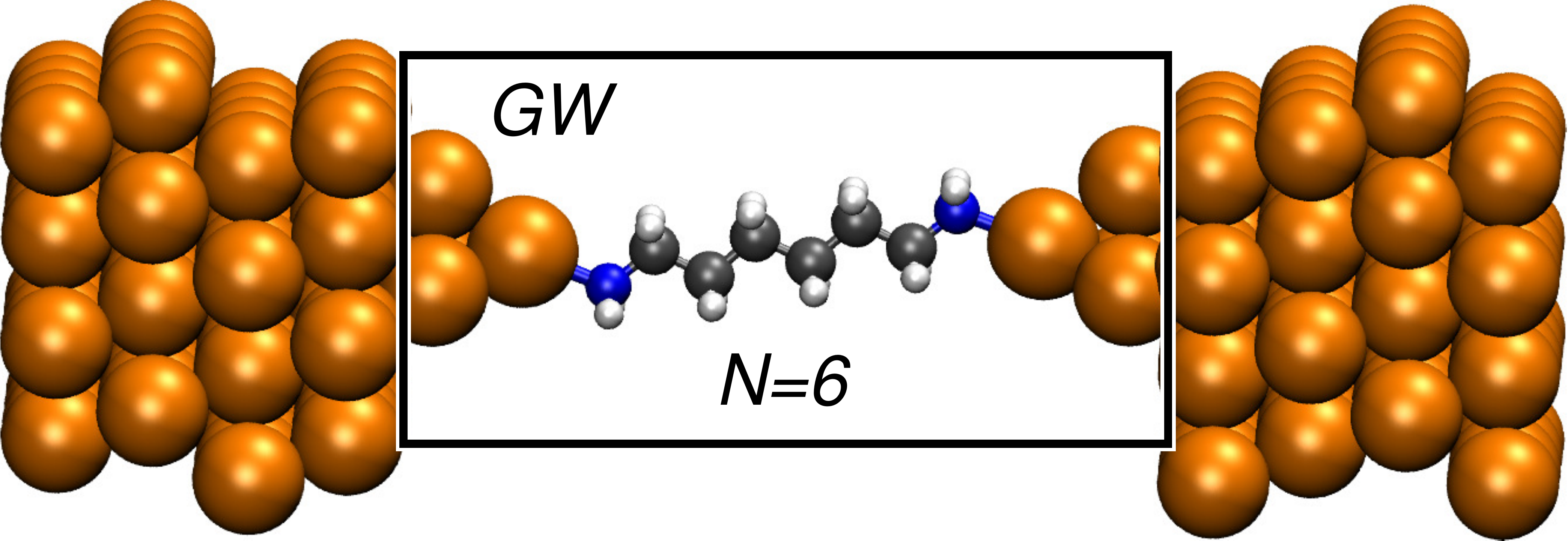}        
\caption{Supercell used to model the gold/alkanediamine junctions. Similar supercells have been used for $N=2$ and 4 (not shown) and key structural parameters can be found here \cite{note}.
The black box indicates the extended molecule region within which the GW self-energy is evaluated. The inclusion of the Au tips in the extended molecule is sufficient to include screening effects from the electrodes on the molecular energy levels.}
\label{fig:figure1}
\end{figure}

  All transport calculations where performed using the method
  described in detail in Ref. \cite{strange}. In brief we
  employ a basis set of numerical atomic orbitals corresponding to
  double-zeta plus polarization (DZP) for the Au atoms and double-zeta
  (DZ) for the atoms of the molecules. We use rather diffuse basis
  functions with a confinement energy shift of 0.01~eV. This ensures
  that the calculated work function of Au(111) and the Kohn-Sham
  energy levels of the molecular junction are within 0.1~eV of those
  obtained from accurate grid calculations \cite{strange}.  The
  transmission function is obtained from the Meir-Wingreen
  transmission formula \cite{ref30,nonorthogonal}
  \begin{equation}\label{eq.eq1}
 T(\varepsilon) = \text{Tr}[G^r(\varepsilon)\Gamma_L(\varepsilon)G^a(\varepsilon) \Gamma_R(\varepsilon)].
 \end{equation}
The retarded Green function of the extended molecule is calculated from
\begin{eqnarray}\nonumber
G^r(E)=[(E+i\eta) S - H_0 + V_{xc} - \Delta
V_{\text{H}}[G]\\\label{eq.gr}
-\Sigma_{L}(E)-\Sigma_{R}(E)-\Sigma_{xc}[G](E)]^{-1}
\end{eqnarray}
Here $S$, $H_0$, and $V_{xc} $ are the overlap matrix, Kohn-Sham
Hamiltonian and the local xc-potential in the atomic orbital basis,
respectively. $\eta$ is a positive infinitesimal.

The lead self-energy, $\Sigma_{L/R}$ , incorporates the coupling to
the left and right electrodes and is obtained using standard
techniques \cite{wannier_thygesen}. The term $\Delta V_{H}$ gives
the change in the Hartree potential relative to the DFT Hartree
potential already contained in $H_0$. Finally, the last term is the
many-body xc self-energy which in this work is either the bare
exchange potential, $V_x$, corresponding to Hartree-Fock or the GW
self-energy. As indicated both the Hartree potential and the xc
self-energy depend on the Green function. The latter is evaluated
fully self-consistently using a simple linear mixing of the Green
functions. The GW self-energy is evaluated for the extended molecule
(indicated by the box in Figure \ref{fig:figure1}). However, only the
part corresponding to the molecule is used while the remaining part is
replaced by the DFT xc-potential. This is done to include non-local
correlation (image charge) effects from the electrodes in the GW
self-energy of the molecule while preserving a consistent description
of all metal atoms at the DFT level. We have verified that the
calculations are converged with respect to the size of the extended
molecules, see Ref. \cite{strange} for more details. We
represent all energy dependent quantities in Equation (\ref{eq.gr}) on
a large energy grid ranging from $-200$~eV to $200$~eV with an energy
grid spacing of 0.01~eV.
\end{section}

\begin{section}{Results and discussion}
\begin{subsection}{Energy level alignment}
  The alignment of the molecular energy levels relative to the
  electrode Fermi level is of great importance for the transport
  properties of molecular junctions and seems to be the dominating effect at low bias
  voltage. At higher bias voltages many body calculations on small
  model systems suggest that electron correlations induce additional
  shifting and  broadening of the molecular levels which can also affect the transport
  properties \cite{thygesen_prl}.  Here we focus on the low bias regime
  and postpone the finite bias effects to a later study.

  The molecular orbitals (MOs) of the alkanediamine chains comprise
  states which are delocalized over the carbon backbone and states
  which are localized on the NH$_2$ end group. We shall consider the
  highest occupied molecular orbital (HOMO) and HOMO-2 as 
  representatives for the two classes of states,
  see Figure \ref{fig:figure2}(a). We note that the HOMO-1 is similar to
  the HOMO with slightly lower energy given by the coupling of the two
  end groups across the wire.  In Table \ref{tab:levels_gas} and
  \ref{tab:levels_junction} we list the energy of the HOMO and
  HOMO-2 calculated with DFT-PBE, Hartree-Fock (HF), and GW for the
  molecules in the gas-phase and junction, respectively.
\begin{table}
\caption{Calculated HOMO and HOMO-2 energies aligned to the vacuum level and in units of eV.}
\label{tab:levels_gas}                                  
\begin{tabular}{c l c c c}
\hline\hline
Method& Orbital & $N=2$&$N=4$&$N=6$\\
 \hline
DFT-PBE & HOMO    &     -4.9    &   -5.1     &    -5.1         \\ 
                  & HOMO-2 &    -8.5    &   -8.2     &    -8.0        \\  
    HF         & HOMO    &   -10.2   &   -10.5   &   -10.5        \\ 
                  & HOMO-2 &   -13.3   &   -12.9   &   -12.8             \\ 
GW           & HOMO     &    -8.5    &   -8.6      &    -8.6        \\ 
                 & HOMO-2  &  -12.2    &   -11.8   &    -11.6       \\ 
\hline\hline
\end{tabular}
\end{table}
\begin{table}
\caption{Calculated HOMO and HOMO-2 energies 
in the junction relative to the electrode Fermi level. }
\label{tab:levels_junction}
\begin{tabular}{c l c c c c}
\hline\hline
Method & Orbital & $N=2$ & $N=4$ & $N=6$ \\
 \hline
 DFT-PBE  & HOMO    &   -4.3   &   -4.2    &   -4.4 \\
                    & HOMO-2&   -6.5   &   -5.9    &   -5.7 \\
 HF             & HOMO    &   -8.1   &   -8.1    &    -8.3 \\
                   & HOMO-2& -10.4  &   -9.9    &    -9.7 \\
 GW           & HOMO   &   -4.8   &   -4.9    &    -5.2 \\
                  & HOMO-2&   -8.4    &  -8.2    &    -8.2 \\
 \hline\hline
\end{tabular}
\end{table}
  In the gas-phase, all three methods predict the HOMO energy to be
  almost independent of molecular length. This is clearly due to its
  end group localized character.  In contrast, the energy of the
  HOMO-2 level shifts upward in energy as the molecular length
  increases.  This reflects its extended nature and can be interpreted
  as a band discretization effect.  To the best of our knowledge no
  experimental results exists for the ionization potential of
  alkanediamine molecules.  However, for the closely related butane
  molecule ($N=2$ alkane with CH$_3$ end groups) we obtain a GW
  calculated HOMO energy of $-11.4$~eV in very good agreement with the
  experimental ionization potential of
  $11.2$~eV \cite{exp_butane_homo}. 
  In comparison, the DFT-PBE HOMO
  energy is severely underestimated at $-7.9$ eV.  This finding agrees
  well with previous studies on a larger range of small
  molecules \cite{rostgaard,strange,blase}.

  In the junction, the molecular orbitals, $|\psi_n\rangle$, have been
  obtained by diagonalizing the DFT Hamiltonian corresponding to the
  molecule. The projected density of states (PDOS) of such a state is
  then given by the spectral function, $-1/\pi\text{Im}\langle
  \psi_n|G^r(E)|\psi_n\rangle$, where $G$ is the appropriate Green
  function (calculated with DFT, HF, or GW). The level position is defined as the
  first moment of the PDOS. Figure \ref{fig:figure2}(b) shows the PDOS
  for the HOMO and HOMO-2 for the $6$-alkanediamine junction as
  calculated with DFT-PBE (upper panel) and GW (middle panel). The
  lower panel shows the PDOS obtained from a DFT calculation where the
  molecular levels have been shifted to match the GW levels, i.e.
  after adding to the Kohn-Sham Hamiltonian a generalized scissors operator of the form,
  $\Sigma_{\text{GSO}}^{\text{DFT}}=\sum_n
  (\varepsilon_n^{\text{QP}}-\varepsilon^{\text{DFT}}_n)
  |\psi_n\rangle\langle\psi_n|$. Here
  $\varepsilon_n^{\text{QP}}$ denote the QP energy obtained from the
  GW calculation. We see that the main features of the GW spectral
  function can be well reproduced by the shifted DFT Hamiltonian although small differences remain. A similar conclusion was reached in Ref.
  \cite{strange} for a gold/benzenediamine junction.
\begin{figure}
\includegraphics[width=0.95\linewidth, angle=0]{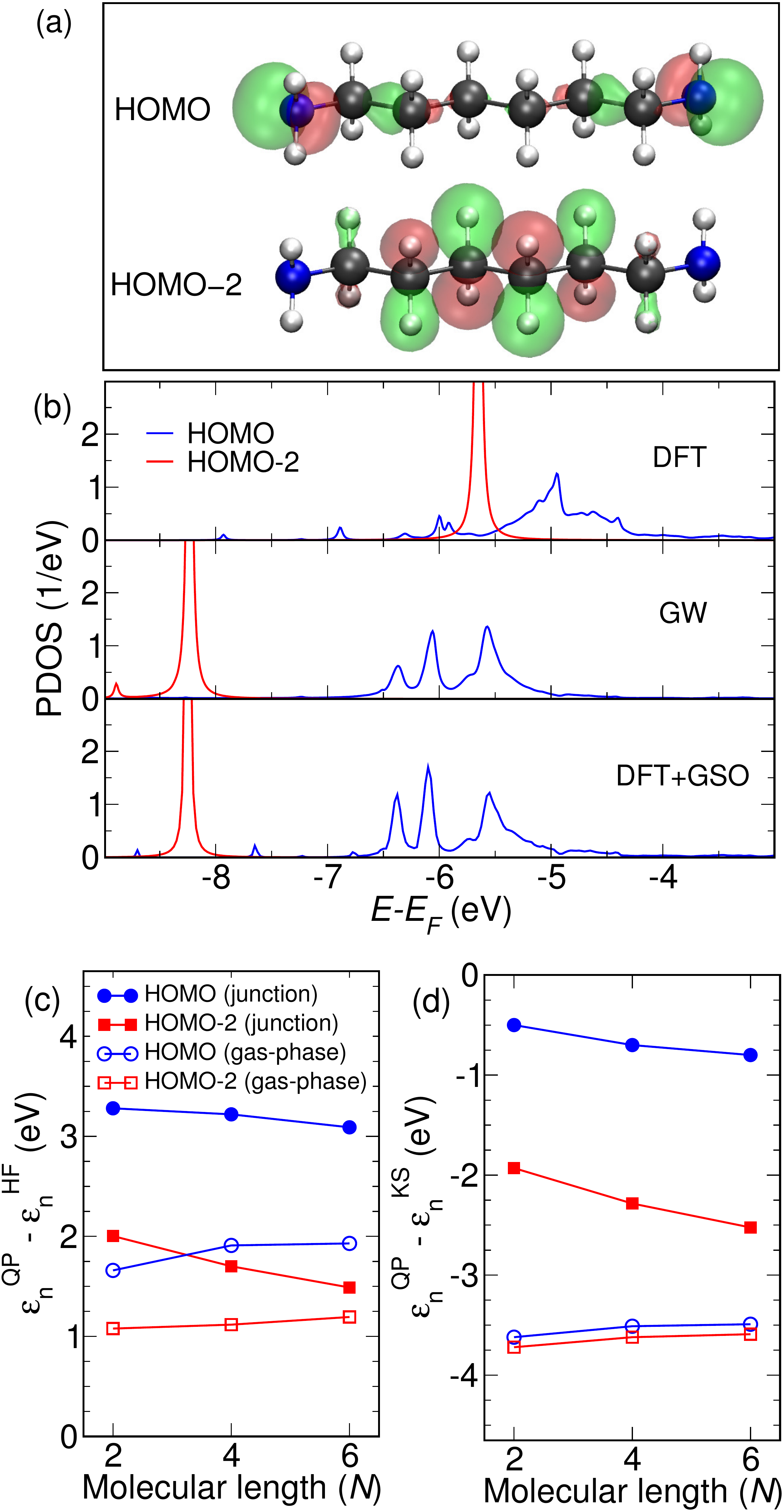}
\caption{(a) Isosurfaces for the HOMO and HOMO-2 orbitals of the $6$-alkanediamine molecule. (b) HOMO and HOMO-2 PDOS in the junction obtained from DFT-PBE (top), GW (middle) and DFT with a generalized scissors operator (bottom). (c) Quasiparticle corrections to the Hartree-Fock levels in the gas-phase (open symbols)  and in the junction (filled symbols) as a function of molecular length $N$.
The HOMO and HOMO-2 are denoted by circles and squares, respectively. (d) Same as (c) but for DFT rather than Hartree-Fock.}
\label{fig:figure2}
\end{figure}
  The molecular orbital energies from a GW calculation include the
  dynamical response of the electron system to an added or removed
  electron through the correlation part of the self-energy.  In
  general correlations tend to shift the filled levels up and the
  empty levels down relative to the bare Hartree-Fock energies. This
  is because the inclusion of screening reduces the energy cost of
  removing/adding electrons to the molecule. When a molecule is
  brought into contact with a metallic junction its environment
  changes from insulating to metallic.  This implies extra screening
  of an added or removed electron which will cause the filled levels
  to shift upwards and the empty levels to shift downwards even more
  than for the isolated molecules, i.e the gap will shrink relative to
  its gas-phase value. It has been shown previously that DFT in
  (semi)local approximations and Hartree-Fock completely misses this
  important effect \cite{neaton,juanma,thygesen_image}.

In Figure \ref{fig:figure2}(c) and (d) we show the QP corrections to the
HF and DFT Kohn-Sham energy levels as function of molecular length.
The results for the HOMO and the HOMO-2 are denoted by circles and
squares. We first notice that the QP corrections are very significant with absolute values reaching almost 4 eV with a pronounced orbital and length dependence. The Hartree-Fock QP corrections are all positive showing that
HF places the occupied levels lower than predicted by GW. This is in
contrast to the corrections to the DFT levels which are all negative
in agreement with the well known underestimation of ionization
potentials by DFT. In contrast to
Hartree-Fock the Kohn-Sham QP corrections are smaller for molecules in
the junction rather than in the gas-phase.  In fact the DFT HOMO level
position is relatively close the GW level position and only lie 0.5-0.8~eV
higher.  The fact that the DFT-PBE description of molecular energy
levels is much better in the junction than in the gas-phase agrees
with previous findings \cite{juanma,strange,ref3}, and can be
explained from the origin of the PBE functional in the homogenous
electron gas \cite{rohlfing}.

It is instructive to consider the shift in the molecular energy levels
due to correlation effects coming from the metal electrodes. In simple terms this corresponds to the shift induced by image charge effects. In order to isolate the
part of the correlation energy originating from the metallic
electrodes we define the quantity,
\begin{equation}
\Delta\varepsilon_{\text{corr},n}=(\varepsilon^\text{QP}_n-\varepsilon^\text{HF}_n)_\text{junction}-(\varepsilon^\text{QP}_n-\varepsilon^\text{HF}_n)_\text{gas-phase}
\end{equation}
which is shown in Figure \ref{fig:figure3}. 
\begin{figure}
\includegraphics[width=0.95\linewidth, angle=0]{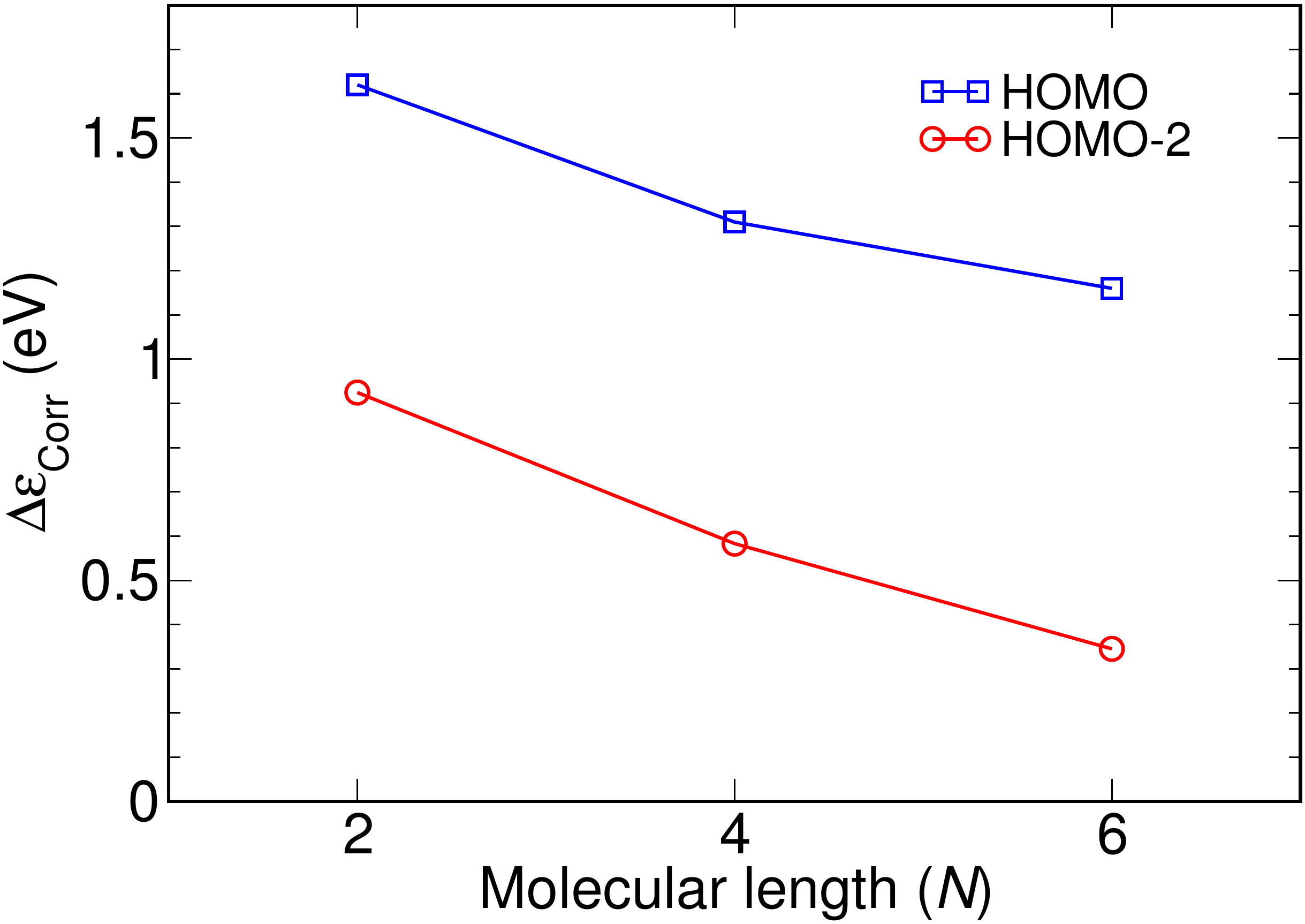}
\caption{Change in the correlation energy of the HOMO and HOMO-2 energy levels when the
molecules are taken from the gas-phase into the junction. This value represents the shift of molecular levels due to the enhanced screening provided by the metallic electrodes (an image charge effect in simple terms).}
\label{fig:figure3} 
\end{figure}
The result can
be understood qualitatively by considering a classical model to account for
the screening effect of the electrodes.  Classically a charge
distribution between two perfect metals will experience an image
potential. For example, the image potential for a point charge halfway
between two metal surfaces separated by a distance $L$ is $\approx
10.0 / L$ (eV \AA)~\cite{stm_book}.  This predicts a $\propto
1/L$ dependence for the image charge effect. The Au tip atoms in our
simulations are about $8$~\AA~apart for the $2$-alkadiamine junction
giving a rough estimate of $1.3$~eV for the image charge effect in
qualitative agreement with the GW calculations. The HOMO experiences a
larger image charge effect than the HOMO-2 which can be understood
from the fact that its charge density is located closer to the
metallic surfaces. In the limit of an infinitely long wire the HOMO-2
will be spread out over the entire molecule and the image charge
effect should vanish. On the other hand in this limit the HOMO will
stay localized near the surface and therefore approach a non-zero
constant image charge potential. If we model the HOMO charge density
as a point charge of half an electron on each of the amine groups we
can estimate this limiting constant to be $3.6 / (2 d)$ (eV \AA),
where $d$ is the distance to the nearest metal surface. Taking $d$ to
be around the Au-N bond length (2.34 \AA) gives a limiting value
estimate of $0.8$ eV. Again, this seems to be in qualitative agreement
with our GW findings.

Finally, we discuss the coverage dependence of the energy level
position for alkanediamine-Au junctions.  It was shown in Ref.
 \cite{wang_prb_2008} that the DFT level position for amine
linked molecules is strongly dependent on coverage. In contrast to the
screening (image charge) effects discussed above, which appear in the
correlation part of the self-energy, this is a purely electrostatic
effect resulting from the localized surface dipoles formed at the
Au-NH$_2$ bond. To investigate the dependence of the energy levels on
coverage for our junctions we have performed DFT calculations for a
range of transverse supercell dimensions for the geometry shown in Figure \ref{fig:figure1}. 
In Figure \ref{fig:figure4} we show the PDOS of a
methylene unit in the central part of the molecule for transverse
super cells with $2\times 2$ to $8\times 8$ surface atoms. 
\begin{figure}
\includegraphics[width=0.95\linewidth, angle=0]{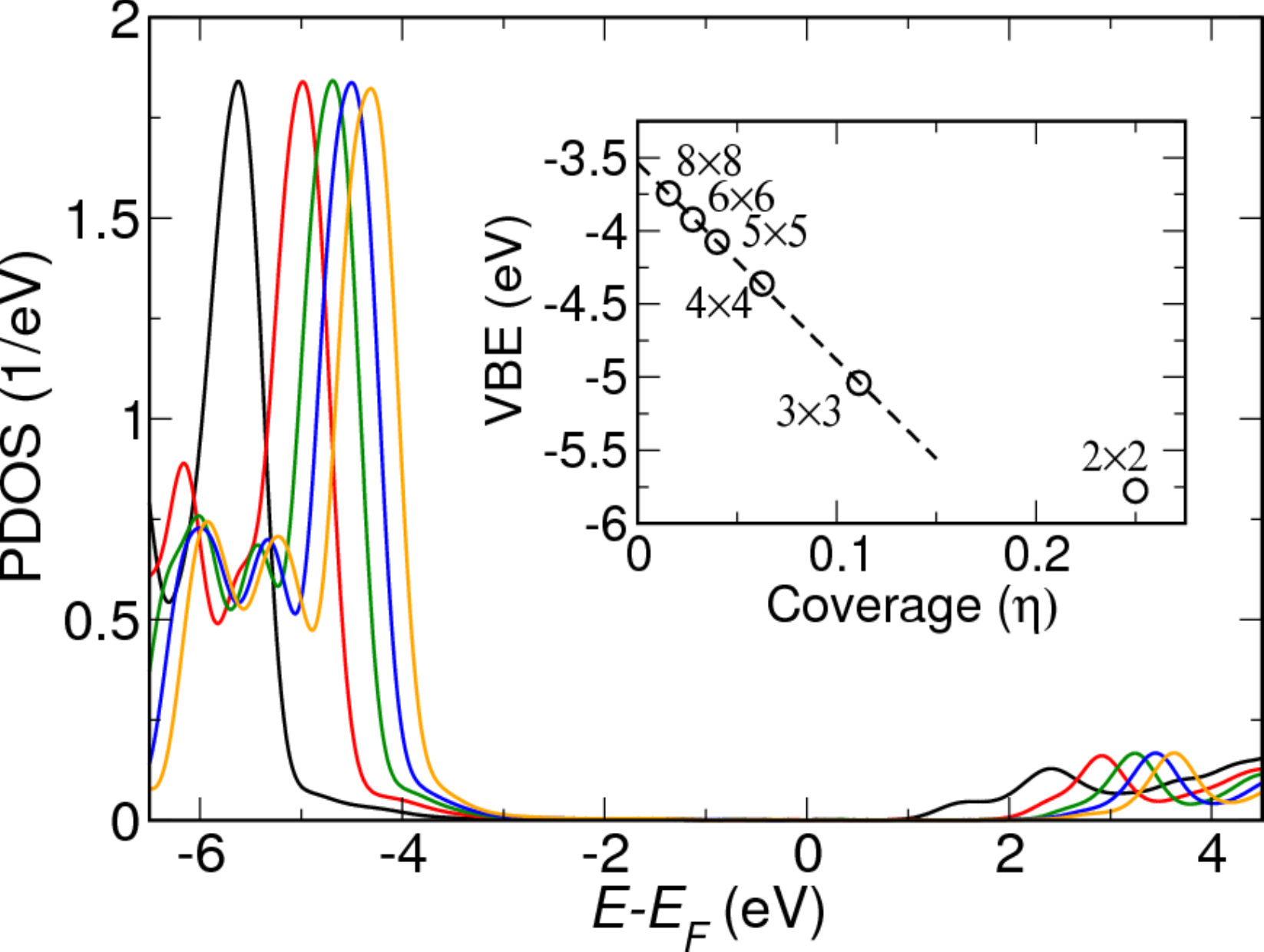}
\caption{The molecular valence band edge (or HOMO) as a function of coverage for an $N=6$ alkane. The numbers indicate the number of surface atoms.}
\label{fig:figure4} 
\end{figure}
The PDOS peaks and band edge is shifting up in energy as the transverse
supercell size is increased in agreement with the results of Ref.
 \cite{wang_prb_2008}.  The inset shows the energy shift
obtained from reading off the shift in the PDOS as a function of
coverage defined as $\eta = 1 / N_\text{surface}$, where
$N_\text{surface}$ is the number of surface atoms. For a supercell
dimension of size $3\times3$ and larger, the shift is seen to be
directly proportional with the coverage as expected for a two
dimensional array of dipoles \cite{natan_2007}.  This allow us to to
extract the electrostatic shift corresponding to the single molecule
limit. We find that the electrostatic energy shift when going from a
$4\times 4$ supercell to the single molecule limit is indeed
significant with a value around 1~eV.
\end{subsection}

\begin{subsection}{Transport calculations}
  The transmission function of the 2, 4 -and 6-alkanediamine junction
  geometries were calculated using the GW and the PBE xc potential as
  approximations for $\Sigma_\text{xc}$ in Eq. (\ref{eq.gr}). To
  include the coverage dependence we
  have simulated the low coverage limit $\eta=0^+$ by performing
  calculations using the $4\times 4$ junction (corresponding to
  $\eta=1/16$) with all molecular levels shifted up by 1 eV using a
  simple scissors operator self-energy.

The transmission function calculated using GW is shown in
Figure \ref{fig:figure5} on a logarithmic scale.  The transmission functions for
different molecular lengths have very similar shapes in the important
region near the Fermi level $E_F$, however, the magnitude is
increasingly suppressed as function of molecular length.  The
similarity of the transmission functions may at first seem surprising
since we have shown that the position of the molecular energy levels
show some length dependence. In particular the HOMO level was found to
decrease in energy by 0.5 eV when $N$ increases from 2 to 6 (see Table \ref{tab:levels_junction}). 
This shift is indeed visible in transmission function in the range -4.0 to -6.0 eV where the HOMO is
located.  On the other hand the feautures in the transmission
function around the Fermi level are determined by the local electronic structure of the Au tips .
\begin{figure}
\includegraphics[width=0.95\linewidth, angle=0]{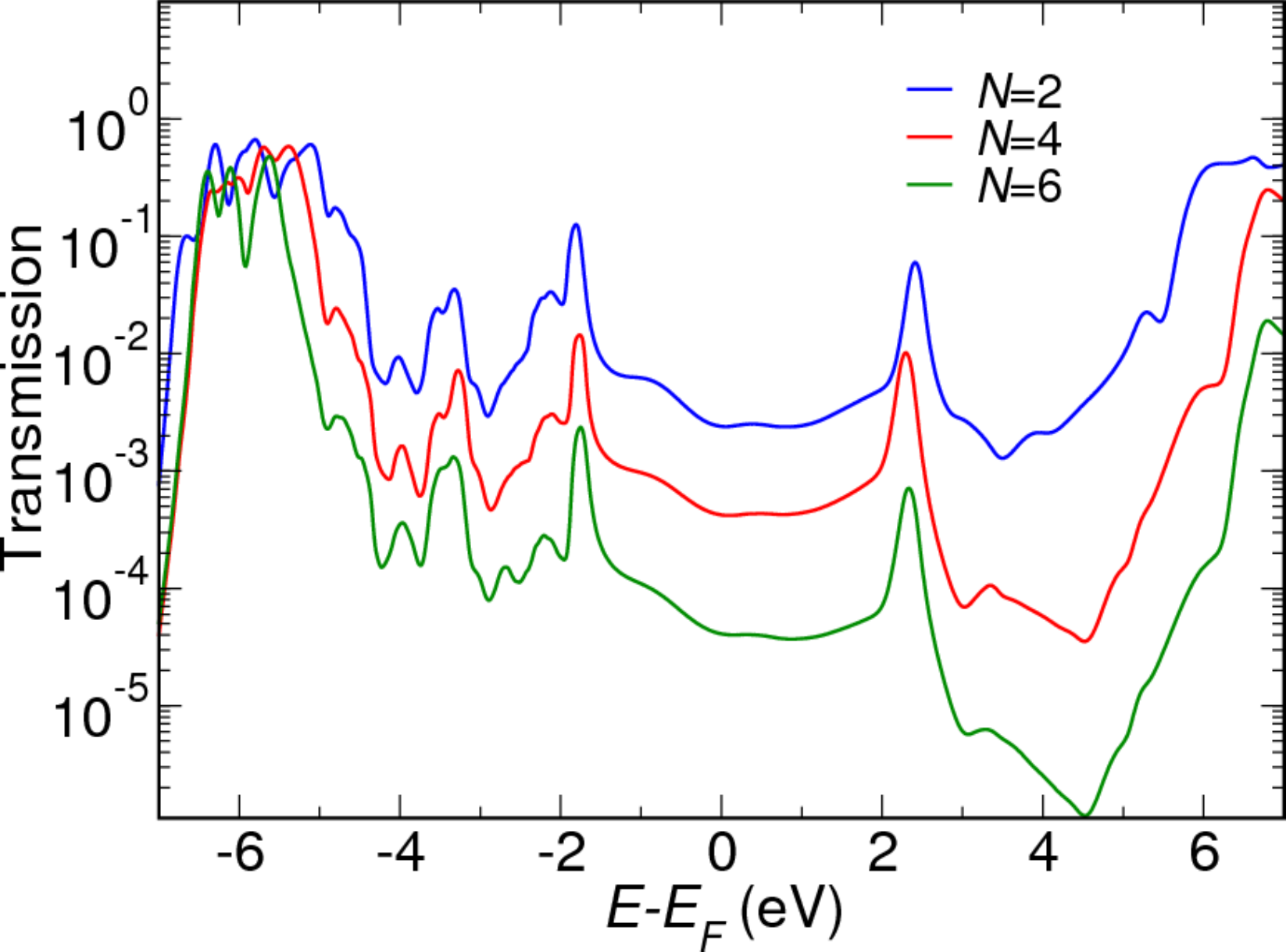}
\caption{The transmission function calculated using GW for a molecular length of $N$=2, 4  and 6.}
\label{fig:figure5}      
\end{figure}

The zero bias conductance is obtained from the transmission function
at the Fermi level, $G=G_0 T(E_F)$ where  $G_0=2e^2/h$ is the unit of quantum conductance. 
The zero bias conductance is plotted in Figure \ref{fig:figure6} as function of molecular length. We have also included the DFT results for comparison.  The dashed lines show the best fits
to the exponential form $G_N=G_c \exp(-\beta N)$. 
\begin{figure}
\includegraphics[width=0.95\linewidth, angle=0]{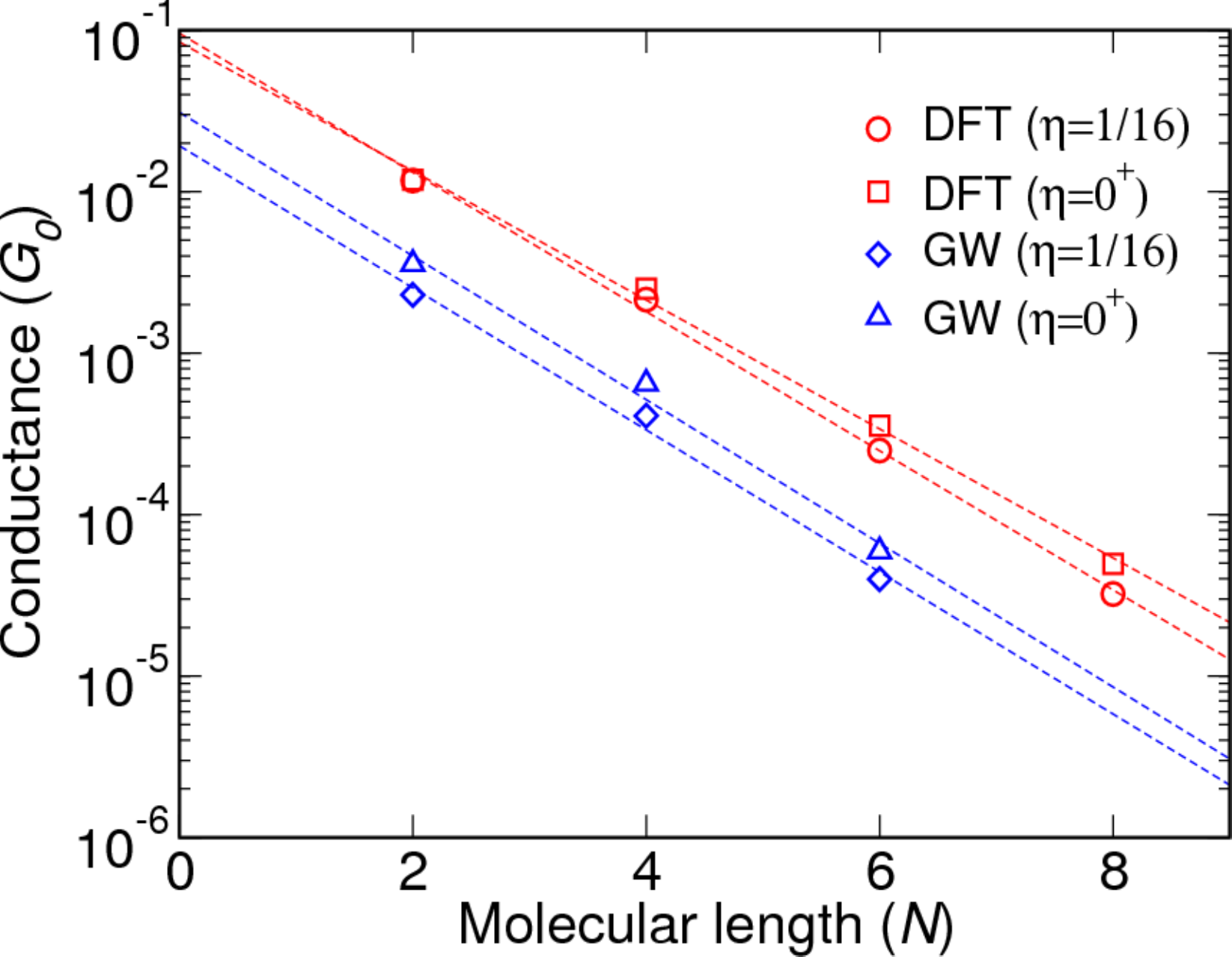}
 \caption{Calculated conductance plotted as a function of the molecular length for coverages corresponding to $4\times 4$ Au atoms per molecule ($\eta=1/16$) and extrapolated to the single-molecule limit ($\eta=0$).}
 \label{fig:figure6} 
\end{figure}
The values for $\beta$ and $G_c$ corresponding to the single-molecule limit ($\eta=0$) and $4\times4$ Au atoms per molecule ($\eta=1/16$) are reported in Table \ref{tab:beta} together with experimental values. 
\begin{table}
\begin{tabular} {c c c c c c}
\hline\hline
             & DFT  &             & GW       &              & Exp.\\
$\eta$  & 1/16 & 0 & $1/16$ & 0  & $-$ \\
\hline          
$G_C$ & 0.11 & 0.10   & 0.02     & 0.04      & $0.02^a$, $0.035^c$  \\
$\beta$& 0.99  & 0.92   & 1.01    & 1.02       & 0.97\footnote{Ref. \cite{hybertsen}}, 0.93\footnote{Ref. \cite{park_jacs_2007}}, 0.91\footnote{Ref. \cite{venkataraman}} \\ 
\hline\hline
\end{tabular}
\caption{Calculated contact conductance $G_C$  
in units of $G_0$ and exponential decay constant $\beta$ per carbon atom.
$\eta$ denotes the coverage.}
\label{tab:beta}
\end{table}
The rather weak effect of coverage on the conduction properties is in agreement with
the findings in Ref. \cite{feng_jpcc_2009} where a
4-alkanediamine Au junction with $3\times3$ and $4\times 4$ surfaces
was considered. Our DFT results are in reasonable agreement with
previous DFT studies showing decay factors in the range $0.83-1.01$
and conductance resistances in the range
$(0.09-0.28)G_0$ \cite{wohlthat_cpl_2008, sheng_jcp_2009,
  mcdermott_jpcc_2009}. While the $\beta$ values obtained with GW are rather close to the DFT calculated ones, the contact conductance is reduced by a
factor of $3-5$ depending on the coverage. This is a direct result of the molecular levels lying further away from $E_F$ (by 0.5-2.5 eV, see Figure \ref{fig:figure2}(d)) in GW compared to DFT. 
\end{subsection}
\end{section}

\begin{section}{Conclusion}
  We have unraveled the important role of exchange-correlation effects
  for the energy level alignment and low-bias conductance of
  gold/alkanediamine molecular junctions. Based on many-body GW
  calculations we found that the origin of the overestimation of the
  contact conductance, $G_c$, by standard DFT is due to wrong energy
  level alignment in the junction. The absence of self-interaction and the inclusion of image charge screening effect through the GW self-energy improves the desription of the energy levels and yield values
  for $G_c$ and the decay constant $\beta$ in good agreement with
  experiments. The quasiparticle corrections to the DFT energy levels
  showed a significant orbital dependence ranging from $-0.5$ eV to
  $-2.5$ eV due to the different shape and localization of the
  molecular orbitals. Our results demonstrate that quantitatively
  accurate calculations of conductance from first-principles are
  feasible, although computationally demanding. 
\end{section}

\begin{section}{Acknowledgments}
The authors acknowledge support from the Lundbeck Foundation's Center for Atomic-scale Materials Design (CAMD) and The Danish Center for Scientific Computing.
\end{section}
\bibliography{paper}

\begin{thebibliography}{61}
\expandafter\ifx\csname natexlab\endcsname\relax\def\natexlab#1{#1}\fi
\expandafter\ifx\csname bibnamefont\endcsname\relax
  \def\bibnamefont#1{#1}\fi
\expandafter\ifx\csname bibfnamefont\endcsname\relax
  \def\bibfnamefont#1{#1}\fi
\expandafter\ifx\csname citenamefont\endcsname\relax
  \def\citenamefont#1{#1}\fi
\expandafter\ifx\csname url\endcsname\relax
  \def\url#1{\texttt{#1}}\fi
\expandafter\ifx\csname urlprefix\endcsname\relax\def\urlprefix{URL }\fi
\providecommand{\bibinfo}[2]{#2}
\providecommand{\eprint}[2][]{\url{#2}}

\bibitem[{\citenamefont{Venkataraman et~al.}(2006)\citenamefont{Venkataraman,
  Klare, Tam, Nuckolls, Hybertsen, and Steigerwald}}]{venkataraman}
\bibinfo{author}{\bibfnamefont{L.}~\bibnamefont{Venkataraman}},
  \bibinfo{author}{\bibfnamefont{J.}~\bibnamefont{Klare}},
  \bibinfo{author}{\bibfnamefont{I.}~\bibnamefont{Tam}},
  \bibinfo{author}{\bibfnamefont{C.}~\bibnamefont{Nuckolls}},
  \bibinfo{author}{\bibfnamefont{M.}~\bibnamefont{Hybertsen}},
  \bibnamefont{and}
  \bibinfo{author}{\bibfnamefont{M.}~\bibnamefont{Steigerwald}},
  \bibinfo{journal}{Nano Lett.} \textbf{\bibinfo{volume}{6}},
  \bibinfo{pages}{458} (\bibinfo{year}{2006}).

\bibitem[{\citenamefont{Kristensen et~al.}(2008)\citenamefont{Kristensen,
  Mowbray, Thygesen, and Jacobsen}}]{iben}
\bibinfo{author}{\bibfnamefont{I.~S.} \bibnamefont{Kristensen}},
  \bibinfo{author}{\bibfnamefont{D.~J.} \bibnamefont{Mowbray}},
  \bibinfo{author}{\bibfnamefont{K.~S.} \bibnamefont{Thygesen}},
  \bibnamefont{and} \bibinfo{author}{\bibfnamefont{K.~W.}
  \bibnamefont{Jacobsen}}, \bibinfo{journal}{J. of Phys.: Condens. Matter}
  \textbf{\bibinfo{volume}{20}}, \bibinfo{pages}{374101}
  (\bibinfo{year}{2008}).

\bibitem[{\citenamefont{Tomfohr and Sankey}(2002)}]{tomfohr_prb_2002}
\bibinfo{author}{\bibfnamefont{J.}~\bibnamefont{Tomfohr}} \bibnamefont{and}
  \bibinfo{author}{\bibfnamefont{O.}~\bibnamefont{Sankey}},
  \bibinfo{journal}{Phys. Rev. B} \textbf{\bibinfo{volume}{65}},
  \bibinfo{pages}{245105} (\bibinfo{year}{2002}).

\bibitem[{\citenamefont{Sen and Kaun}(2010)}]{sen_acsnano_2010}
\bibinfo{author}{\bibfnamefont{A.}~\bibnamefont{Sen}} \bibnamefont{and}
  \bibinfo{author}{\bibfnamefont{C.-C.} \bibnamefont{Kaun}},
  \bibinfo{journal}{ACS Nano} \textbf{\bibinfo{volume}{4}},
  \bibinfo{pages}{6404} (\bibinfo{year}{2010}).

\bibitem[{\citenamefont{Kaun and Guo}(2003)}]{kaun_nanolett_2003}
\bibinfo{author}{\bibfnamefont{C.}~\bibnamefont{Kaun}} \bibnamefont{and}
  \bibinfo{author}{\bibfnamefont{H.}~\bibnamefont{Guo}}, \bibinfo{journal}{Nano
  Lett.} \textbf{\bibinfo{volume}{3}}, \bibinfo{pages}{1521}
  (\bibinfo{year}{2003}).

\bibitem[{\citenamefont{Muller}(2006)}]{muller_prb_2006}
\bibinfo{author}{\bibfnamefont{K.}~\bibnamefont{Muller}},
  \bibinfo{journal}{Phys. Rev. B} \textbf{\bibinfo{volume}{73}},
  \bibinfo{pages}{045403} (\bibinfo{year}{2006}).

\bibitem[{\citenamefont{Paulsson et~al.}(2009)\citenamefont{Paulsson, Krag,
  Frederiksen, and Brandbyge}}]{magnus}
\bibinfo{author}{\bibfnamefont{M.}~\bibnamefont{Paulsson}},
  \bibinfo{author}{\bibfnamefont{C.}~\bibnamefont{Krag}},
  \bibinfo{author}{\bibfnamefont{T.}~\bibnamefont{Frederiksen}},
  \bibnamefont{and}
  \bibinfo{author}{\bibfnamefont{M.}~\bibnamefont{Brandbyge}},
  \bibinfo{journal}{Nano Lett.} \textbf{\bibinfo{volume}{9}},
  \bibinfo{pages}{117} (\bibinfo{year}{2009}).

\bibitem[{\citenamefont{Wohlthat et~al.}(2008)\citenamefont{Wohlthat, Pauly,
  and Reimers}}]{wohlthat_cpl_2008}
\bibinfo{author}{\bibfnamefont{S.}~\bibnamefont{Wohlthat}},
  \bibinfo{author}{\bibfnamefont{F.}~\bibnamefont{Pauly}}, \bibnamefont{and}
  \bibinfo{author}{\bibfnamefont{J.~R.} \bibnamefont{Reimers}},
  \bibinfo{journal}{Chem. Phys. Lett.} \textbf{\bibinfo{volume}{454}},
  \bibinfo{pages}{284} (\bibinfo{year}{2008}).

\bibitem[{\citenamefont{McDermott et~al.}(2009)\citenamefont{McDermott, George,
  Fagas, Greer, and Ratner}}]{mcdermott_jpcc_2009}
\bibinfo{author}{\bibfnamefont{S.}~\bibnamefont{McDermott}},
  \bibinfo{author}{\bibfnamefont{C.~B.} \bibnamefont{George}},
  \bibinfo{author}{\bibfnamefont{G.}~\bibnamefont{Fagas}},
  \bibinfo{author}{\bibfnamefont{J.~C.} \bibnamefont{Greer}}, \bibnamefont{and}
  \bibinfo{author}{\bibfnamefont{M.~A.} \bibnamefont{Ratner}},
  \bibinfo{journal}{J. Phys. Chem. C} \textbf{\bibinfo{volume}{113}},
  \bibinfo{pages}{744} (\bibinfo{year}{2009}).

\bibitem[{\citenamefont{Sheng et~al.}(2009)\citenamefont{Sheng, Li, Ning,
  Zhang, Yang, and Guo}}]{sheng_jcp_2009}
\bibinfo{author}{\bibfnamefont{W.}~\bibnamefont{Sheng}},
  \bibinfo{author}{\bibfnamefont{Z.~Y.} \bibnamefont{Li}},
  \bibinfo{author}{\bibfnamefont{Z.~Y.} \bibnamefont{Ning}},
  \bibinfo{author}{\bibfnamefont{Z.~H.} \bibnamefont{Zhang}},
  \bibinfo{author}{\bibfnamefont{Z.~Q.} \bibnamefont{Yang}}, \bibnamefont{and}
  \bibinfo{author}{\bibfnamefont{H.}~\bibnamefont{Guo}}, \bibinfo{journal}{J.
  Chem. Phys.} \textbf{\bibinfo{volume}{131}}, \bibinfo{pages}{244712}
  (\bibinfo{year}{2009}).

\bibitem[{\citenamefont{Li et~al.}(2008)\citenamefont{Li, Pobelov, Wandlowski,
  Bagrets, Arnold, and Evers}}]{evers}
\bibinfo{author}{\bibfnamefont{C.}~\bibnamefont{Li}},
  \bibinfo{author}{\bibfnamefont{I.}~\bibnamefont{Pobelov}},
  \bibinfo{author}{\bibfnamefont{T.}~\bibnamefont{Wandlowski}},
  \bibinfo{author}{\bibfnamefont{A.}~\bibnamefont{Bagrets}},
  \bibinfo{author}{\bibfnamefont{A.}~\bibnamefont{Arnold}}, \bibnamefont{and}
  \bibinfo{author}{\bibfnamefont{F.}~\bibnamefont{Evers}}, \bibinfo{journal}{J.
  Am. Chem. Soc.} \textbf{\bibinfo{volume}{130}}, \bibinfo{pages}{318}
  (\bibinfo{year}{2008}).

\bibitem[{\citenamefont{Hybertsen et~al.}(2008)\citenamefont{Hybertsen,
  Venkataraman, Klare, CWhalley, Steigerwald, and Nuckolls}}]{hybertsen}
\bibinfo{author}{\bibfnamefont{M.~S.} \bibnamefont{Hybertsen}},
  \bibinfo{author}{\bibfnamefont{L.}~\bibnamefont{Venkataraman}},
  \bibinfo{author}{\bibfnamefont{J.~E.} \bibnamefont{Klare}},
  \bibinfo{author}{\bibfnamefont{A.}~\bibnamefont{CWhalley}},
  \bibinfo{author}{\bibfnamefont{M.~L.} \bibnamefont{Steigerwald}},
  \bibnamefont{and} \bibinfo{author}{\bibfnamefont{C.}~\bibnamefont{Nuckolls}},
  \bibinfo{journal}{J. Phys.:Condens. Matter} \textbf{\bibinfo{volume}{20}},
  \bibinfo{pages}{374115} (\bibinfo{year}{2008}).

\bibitem[{\citenamefont{Quek et~al.}(2007)\citenamefont{Quek, Venkataraman,
  Choi, Louie, Hybertsen, and Neaton}}]{quek}
\bibinfo{author}{\bibfnamefont{S.~Y.} \bibnamefont{Quek}},
  \bibinfo{author}{\bibfnamefont{L.}~\bibnamefont{Venkataraman}},
  \bibinfo{author}{\bibfnamefont{H.~J.} \bibnamefont{Choi}},
  \bibinfo{author}{\bibfnamefont{S.~G.} \bibnamefont{Louie}},
  \bibinfo{author}{\bibfnamefont{M.~S.} \bibnamefont{Hybertsen}},
  \bibnamefont{and} \bibinfo{author}{\bibfnamefont{J.~B.}
  \bibnamefont{Neaton}}, \bibinfo{journal}{Nano Lett.}
  \textbf{\bibinfo{volume}{7}}, \bibinfo{pages}{3477} (\bibinfo{year}{2007}).

\bibitem[{\citenamefont{Mowbray et~al.}(2008)\citenamefont{Mowbray, Jones, and
  Thygesen}}]{mowbray}
\bibinfo{author}{\bibfnamefont{D.~J.} \bibnamefont{Mowbray}},
  \bibinfo{author}{\bibfnamefont{G.}~\bibnamefont{Jones}}, \bibnamefont{and}
  \bibinfo{author}{\bibfnamefont{K.~S.} \bibnamefont{Thygesen}},
  \bibinfo{journal}{J. Chem. Phys.} \textbf{\bibinfo{volume}{128}},
  \bibinfo{pages}{111103} (\bibinfo{year}{2008}).

\bibitem[{\citenamefont{Smit et~al.}(2002)\citenamefont{Smit, Noat, Untiedt,
  Lang, van Hemert, and van Ruitenbeek}}]{smit02}
\bibinfo{author}{\bibfnamefont{R.}~\bibnamefont{Smit}},
  \bibinfo{author}{\bibfnamefont{Y.}~\bibnamefont{Noat}},
  \bibinfo{author}{\bibfnamefont{C.}~\bibnamefont{Untiedt}},
  \bibinfo{author}{\bibfnamefont{N.}~\bibnamefont{Lang}},
  \bibinfo{author}{\bibfnamefont{M.}~\bibnamefont{van Hemert}},
  \bibnamefont{and} \bibinfo{author}{\bibfnamefont{J.}~\bibnamefont{van
  Ruitenbeek}}, \bibinfo{journal}{Nature (London)}
  \textbf{\bibinfo{volume}{419}}, \bibinfo{pages}{906} (\bibinfo{year}{2002}).

\bibitem[{\citenamefont{Djukic et~al.}(2005)\citenamefont{Djukic, Thygesen,
  Untiedt, Smit, Jacobsen, and van Ruitenbeek}}]{thygesen_H2}
\bibinfo{author}{\bibfnamefont{D.}~\bibnamefont{Djukic}},
  \bibinfo{author}{\bibfnamefont{K.~S.} \bibnamefont{Thygesen}},
  \bibinfo{author}{\bibfnamefont{C.}~\bibnamefont{Untiedt}},
  \bibinfo{author}{\bibfnamefont{R.}~\bibnamefont{Smit}},
  \bibinfo{author}{\bibfnamefont{K.}~\bibnamefont{Jacobsen}}, \bibnamefont{and}
  \bibinfo{author}{\bibfnamefont{J.}~\bibnamefont{van Ruitenbeek}},
  \bibinfo{journal}{Phys. Rev. B} \textbf{\bibinfo{volume}{71}},
  \bibinfo{pages}{161402} (\bibinfo{year}{2005}).

\bibitem[{\citenamefont{Untiedt et~al.}(2004)\citenamefont{Untiedt, Dekker,
  Djukic, and van Ruitenbeek}}]{untiedt04}
\bibinfo{author}{\bibfnamefont{C.}~\bibnamefont{Untiedt}},
  \bibinfo{author}{\bibfnamefont{D.}~\bibnamefont{Dekker}},
  \bibinfo{author}{\bibfnamefont{D.}~\bibnamefont{Djukic}}, \bibnamefont{and}
  \bibinfo{author}{\bibfnamefont{J.}~\bibnamefont{van Ruitenbeek}},
  \bibinfo{journal}{Phys. Rev. B} \textbf{\bibinfo{volume}{69}},
  \bibinfo{pages}{081401} (\bibinfo{year}{2004}).

\bibitem[{\citenamefont{Strange et~al.}(2006)\citenamefont{Strange, Thygesen,
  and Jacobsen}}]{strange_CO}
\bibinfo{author}{\bibfnamefont{M.}~\bibnamefont{Strange}},
  \bibinfo{author}{\bibfnamefont{K.~S.} \bibnamefont{Thygesen}},
  \bibnamefont{and} \bibinfo{author}{\bibfnamefont{K.~W.}
  \bibnamefont{Jacobsen}}, \bibinfo{journal}{Phys. Rev. B}
  \textbf{\bibinfo{volume}{73}}, \bibinfo{pages}{125424}
  (\bibinfo{year}{2006}).

\bibitem[{\citenamefont{Hedin}(1965)}]{hedin}
\bibinfo{author}{\bibfnamefont{L.}~\bibnamefont{Hedin}},
  \bibinfo{journal}{Phys. Rev.} \textbf{\bibinfo{volume}{139}},
  \bibinfo{pages}{A796} (\bibinfo{year}{1965}).

\bibitem[{\citenamefont{Strange et~al.}(2011)\citenamefont{Strange, Rostgaard,
  Hakkinen, and Thygesen}}]{strange}
\bibinfo{author}{\bibfnamefont{M.}~\bibnamefont{Strange}},
  \bibinfo{author}{\bibfnamefont{C.}~\bibnamefont{Rostgaard}},
  \bibinfo{author}{\bibfnamefont{H.}~\bibnamefont{Hakkinen}}, \bibnamefont{and}
  \bibinfo{author}{\bibfnamefont{K.~S.} \bibnamefont{Thygesen}},
  \bibinfo{journal}{Phys. Rev. B} \textbf{\bibinfo{volume}{83}},
  \bibinfo{pages}{115108} (\bibinfo{year}{2011}).

\bibitem[{\citenamefont{Neaton et~al.}(2006)\citenamefont{Neaton, Hybertsen,
  and Louie}}]{neaton}
\bibinfo{author}{\bibfnamefont{J.~B.} \bibnamefont{Neaton}},
  \bibinfo{author}{\bibfnamefont{M.~S.} \bibnamefont{Hybertsen}},
  \bibnamefont{and} \bibinfo{author}{\bibfnamefont{S.~G.} \bibnamefont{Louie}},
  \bibinfo{journal}{Phys. Rev. Lett.} \textbf{\bibinfo{volume}{97}},
  \bibinfo{pages}{216405} (\bibinfo{year}{2006}).

\bibitem[{\citenamefont{Garcia-Lastra et~al.}(2009)\citenamefont{Garcia-Lastra,
  Rostgaard, Rubio, and Thygesen}}]{juanma}
\bibinfo{author}{\bibfnamefont{J.~M.} \bibnamefont{Garcia-Lastra}},
  \bibinfo{author}{\bibfnamefont{C.}~\bibnamefont{Rostgaard}},
  \bibinfo{author}{\bibfnamefont{A.}~\bibnamefont{Rubio}}, \bibnamefont{and}
  \bibinfo{author}{\bibfnamefont{K.~S.} \bibnamefont{Thygesen}},
  \bibinfo{journal}{Phys. Rev. B} \textbf{\bibinfo{volume}{80}},
  \bibinfo{pages}{245427} (\bibinfo{year}{2009}).

\bibitem[{\citenamefont{Thygesen and Rubio}(2009)}]{thygesen_image}
\bibinfo{author}{\bibfnamefont{K.~S.} \bibnamefont{Thygesen}} \bibnamefont{and}
  \bibinfo{author}{\bibfnamefont{A.}~\bibnamefont{Rubio}},
  \bibinfo{journal}{Phys. Rev. Lett.} \textbf{\bibinfo{volume}{102}},
  \bibinfo{pages}{046802} (\bibinfo{year}{2009}).

\bibitem[{\citenamefont{Rostgaard et~al.}(2010)\citenamefont{Rostgaard,
  Jacobsen, and Thygesen}}]{rostgaard}
\bibinfo{author}{\bibfnamefont{C.}~\bibnamefont{Rostgaard}},
  \bibinfo{author}{\bibfnamefont{K.~W.} \bibnamefont{Jacobsen}},
  \bibnamefont{and} \bibinfo{author}{\bibfnamefont{K.~S.}
  \bibnamefont{Thygesen}}, \bibinfo{journal}{Phys. Rev. B}
  \textbf{\bibinfo{volume}{81}}, \bibinfo{pages}{085103}
  (\bibinfo{year}{2010}).

\bibitem[{\citenamefont{Blase et~al.}(2011)\citenamefont{Blase, Attaccalite,
  and Olevano}}]{blase}
\bibinfo{author}{\bibfnamefont{X.}~\bibnamefont{Blase}},
  \bibinfo{author}{\bibfnamefont{C.}~\bibnamefont{Attaccalite}},
  \bibnamefont{and} \bibinfo{author}{\bibfnamefont{V.}~\bibnamefont{Olevano}},
  \bibinfo{journal}{Phys. Rev. B} \textbf{\bibinfo{volume}{83}},
  \bibinfo{pages}{115103} (\bibinfo{year}{2011}).

\bibitem[{\citenamefont{Hybertsen and Louie}(1986)}]{hybertsen2}
\bibinfo{author}{\bibfnamefont{M.}~\bibnamefont{Hybertsen}} \bibnamefont{and}
  \bibinfo{author}{\bibfnamefont{S.}~\bibnamefont{Louie}},
  \bibinfo{journal}{Phys. Rev. B} \textbf{\bibinfo{volume}{34}},
  \bibinfo{pages}{5390} (\bibinfo{year}{1986}).

\bibitem[{\citenamefont{Holm and von Barth}(1998)}]{barth}
\bibinfo{author}{\bibfnamefont{B.}~\bibnamefont{Holm}} \bibnamefont{and}
  \bibinfo{author}{\bibfnamefont{U.}~\bibnamefont{von Barth}},
  \bibinfo{journal}{Phys. Rev. B} \textbf{\bibinfo{volume}{57}},
  \bibinfo{pages}{2108} (\bibinfo{year}{1998}).

\bibitem[{\citenamefont{Xu and Tao}(2003)}]{xu_science_2003}
\bibinfo{author}{\bibfnamefont{B.}~\bibnamefont{Xu}} \bibnamefont{and}
  \bibinfo{author}{\bibfnamefont{N.}~\bibnamefont{Tao}},
  \bibinfo{journal}{Science} \textbf{\bibinfo{volume}{301}},
  \bibinfo{pages}{1221} (\bibinfo{year}{2003}).

\bibitem[{\citenamefont{Engelkes et~al.}(2004)\citenamefont{Engelkes, Beebe,
  and Frisbie}}]{engelkes_jacs_2004}
\bibinfo{author}{\bibfnamefont{V.}~\bibnamefont{Engelkes}},
  \bibinfo{author}{\bibfnamefont{J.}~\bibnamefont{Beebe}}, \bibnamefont{and}
  \bibinfo{author}{\bibfnamefont{C.}~\bibnamefont{Frisbie}},
  \bibinfo{journal}{J. Am. Chem. Soc.} \textbf{\bibinfo{volume}{126}},
  \bibinfo{pages}{14287} (\bibinfo{year}{2004}).

\bibitem[{\citenamefont{Chen et~al.}(2006)\citenamefont{Chen, Li, Hihath,
  Huang, and Tao}}]{chen_jacs_2006}
\bibinfo{author}{\bibfnamefont{F.}~\bibnamefont{Chen}},
  \bibinfo{author}{\bibfnamefont{X.}~\bibnamefont{Li}},
  \bibinfo{author}{\bibfnamefont{J.}~\bibnamefont{Hihath}},
  \bibinfo{author}{\bibfnamefont{Z.}~\bibnamefont{Huang}}, \bibnamefont{and}
  \bibinfo{author}{\bibfnamefont{N.}~\bibnamefont{Tao}}, \bibinfo{journal}{J.
  Am. Chem. Soc.} \textbf{\bibinfo{volume}{128}}, \bibinfo{pages}{15874}
  (\bibinfo{year}{2006}).

\bibitem[{\citenamefont{Park et~al.}(2007)\citenamefont{Park, Whalley,
  Kamenetska, Steigerwald, Hybertsen, Nuckolls, and
  Venkataraman}}]{park_jacs_2007}
\bibinfo{author}{\bibfnamefont{Y.~S.} \bibnamefont{Park}},
  \bibinfo{author}{\bibfnamefont{A.~C.} \bibnamefont{Whalley}},
  \bibinfo{author}{\bibfnamefont{M.}~\bibnamefont{Kamenetska}},
  \bibinfo{author}{\bibfnamefont{M.~L.} \bibnamefont{Steigerwald}},
  \bibinfo{author}{\bibfnamefont{M.~S.} \bibnamefont{Hybertsen}},
  \bibinfo{author}{\bibfnamefont{C.}~\bibnamefont{Nuckolls}}, \bibnamefont{and}
  \bibinfo{author}{\bibfnamefont{L.}~\bibnamefont{Venkataraman}},
  \bibinfo{journal}{J. Am. Chem. Soc.} \textbf{\bibinfo{volume}{129}},
  \bibinfo{pages}{15768} (\bibinfo{year}{2007}).

\bibitem[{\citenamefont{Huang et~al.}(2007)\citenamefont{Huang, Chen, Bennett,
  and Tao}}]{huang_jacs_2007}
\bibinfo{author}{\bibfnamefont{Z.}~\bibnamefont{Huang}},
  \bibinfo{author}{\bibfnamefont{F.}~\bibnamefont{Chen}},
  \bibinfo{author}{\bibfnamefont{P.~A.} \bibnamefont{Bennett}},
  \bibnamefont{and} \bibinfo{author}{\bibfnamefont{N.}~\bibnamefont{Tao}},
  \bibinfo{journal}{J. Am. Chem. Soc.} \textbf{\bibinfo{volume}{129}},
  \bibinfo{pages}{13225} (\bibinfo{year}{2007}).

\bibitem[{\citenamefont{Widawsky et~al.}(2009)\citenamefont{Widawsky,
  Kamenetska, Klare, Nuckolls, Steigerwald, Hybertsen, and
  Venkataraman}}]{widawsky_nanotechnology_2009}
\bibinfo{author}{\bibfnamefont{J.~R.} \bibnamefont{Widawsky}},
  \bibinfo{author}{\bibfnamefont{M.}~\bibnamefont{Kamenetska}},
  \bibinfo{author}{\bibfnamefont{J.}~\bibnamefont{Klare}},
  \bibinfo{author}{\bibfnamefont{C.}~\bibnamefont{Nuckolls}},
  \bibinfo{author}{\bibfnamefont{M.~L.} \bibnamefont{Steigerwald}},
  \bibinfo{author}{\bibfnamefont{M.~S.} \bibnamefont{Hybertsen}},
  \bibnamefont{and}
  \bibinfo{author}{\bibfnamefont{L.}~\bibnamefont{Venkataraman}},
  \bibinfo{journal}{Nanotechnology} \textbf{\bibinfo{volume}{20}},
  \bibinfo{pages}{434009} (\bibinfo{year}{2009}).

\bibitem[{\citenamefont{Kamenetska et~al.}(2009)\citenamefont{Kamenetska,
  koentopp, Whalley, Park, Steigerwald, Nuckolls, Hybertsen, and
  Venkataraman}}]{kamenetska_prl_2009}
\bibinfo{author}{\bibfnamefont{M.}~\bibnamefont{Kamenetska}},
  \bibinfo{author}{\bibfnamefont{M.}~\bibnamefont{koentopp}},
  \bibinfo{author}{\bibfnamefont{A.~C.} \bibnamefont{Whalley}},
  \bibinfo{author}{\bibfnamefont{Y.~S.} \bibnamefont{Park}},
  \bibinfo{author}{\bibfnamefont{M.~L.} \bibnamefont{Steigerwald}},
  \bibinfo{author}{\bibfnamefont{C.}~\bibnamefont{Nuckolls}},
  \bibinfo{author}{\bibfnamefont{M.~S.} \bibnamefont{Hybertsen}},
  \bibnamefont{and}
  \bibinfo{author}{\bibfnamefont{L.}~\bibnamefont{Venkataraman}},
  \bibinfo{journal}{Phys. Rev. Lett.} \textbf{\bibinfo{volume}{102}},
  \bibinfo{pages}{126803} (\bibinfo{year}{2009}).

\bibitem[{\citenamefont{Song et~al.}(2007)\citenamefont{Song, Lee, Choi, and
  Lee}}]{song_apl_2007}
\bibinfo{author}{\bibfnamefont{H.}~\bibnamefont{Song}},
  \bibinfo{author}{\bibfnamefont{T.}~\bibnamefont{Lee}},
  \bibinfo{author}{\bibfnamefont{N.-J.} \bibnamefont{Choi}}, \bibnamefont{and}
  \bibinfo{author}{\bibfnamefont{H.}~\bibnamefont{Lee}},
  \bibinfo{journal}{Appl. Phys. Lett.} \textbf{\bibinfo{volume}{91}},
  \bibinfo{pages}{253116} (\bibinfo{year}{2007}).

\bibitem[{\citenamefont{Kim et~al.}(2011)\citenamefont{Kim, Hellmuth, Buerkle,
  Pauly, and Scheer}}]{kim_nano_2011}
\bibinfo{author}{\bibfnamefont{Y.}~\bibnamefont{Kim}},
  \bibinfo{author}{\bibfnamefont{T.~J.} \bibnamefont{Hellmuth}},
  \bibinfo{author}{\bibfnamefont{M.}~\bibnamefont{Buerkle}},
  \bibinfo{author}{\bibfnamefont{F.}~\bibnamefont{Pauly}}, \bibnamefont{and}
  \bibinfo{author}{\bibfnamefont{E.}~\bibnamefont{Scheer}},
  \bibinfo{journal}{ACS Nano} \textbf{\bibinfo{volume}{5}},
  \bibinfo{pages}{4104} (\bibinfo{year}{2011}).

\bibitem[{\citenamefont{Martin et~al.}(2008)\citenamefont{Martin, Ding, van~der
  Zant, and van Ruitenbeek}}]{ruitenbeek}
\bibinfo{author}{\bibfnamefont{C.~A.} \bibnamefont{Martin}},
  \bibinfo{author}{\bibfnamefont{D.}~\bibnamefont{Ding}},
  \bibinfo{author}{\bibfnamefont{H.~S.~J.} \bibnamefont{van~der Zant}},
  \bibnamefont{and} \bibinfo{author}{\bibfnamefont{J.~M.} \bibnamefont{van
  Ruitenbeek}}, \bibinfo{journal}{New J. Phys.} \textbf{\bibinfo{volume}{10}},
  \bibinfo{pages}{065008} (\bibinfo{year}{2008}).

\bibitem[{\citenamefont{Nichols et~al.}(2010)\citenamefont{Nichols, Haiss,
  Higgins, Leary, Martin, and Bethell}}]{nichols_pccp_2010}
\bibinfo{author}{\bibfnamefont{R.~J.} \bibnamefont{Nichols}},
  \bibinfo{author}{\bibfnamefont{W.}~\bibnamefont{Haiss}},
  \bibinfo{author}{\bibfnamefont{S.~J.} \bibnamefont{Higgins}},
  \bibinfo{author}{\bibfnamefont{E.}~\bibnamefont{Leary}},
  \bibinfo{author}{\bibfnamefont{S.}~\bibnamefont{Martin}}, \bibnamefont{and}
  \bibinfo{author}{\bibfnamefont{D.}~\bibnamefont{Bethell}},
  \bibinfo{journal}{Phys. Chem. Chem. Phys.} \textbf{\bibinfo{volume}{12}},
  \bibinfo{pages}{2801} (\bibinfo{year}{2010}).

\bibitem[{\citenamefont{Haiss et~al.}(2009)\citenamefont{Haiss, Martin,
  Scullion, Bouffier, Higgins, and Nichols}}]{haiss_pccp_2009}
\bibinfo{author}{\bibfnamefont{W.}~\bibnamefont{Haiss}},
  \bibinfo{author}{\bibfnamefont{S.}~\bibnamefont{Martin}},
  \bibinfo{author}{\bibfnamefont{L.~E.} \bibnamefont{Scullion}},
  \bibinfo{author}{\bibfnamefont{L.}~\bibnamefont{Bouffier}},
  \bibinfo{author}{\bibfnamefont{S.~J.} \bibnamefont{Higgins}},
  \bibnamefont{and} \bibinfo{author}{\bibfnamefont{R.~J.}
  \bibnamefont{Nichols}}, \bibinfo{journal}{Phys. Chem. Chem. Phys.}
  \textbf{\bibinfo{volume}{11}}, \bibinfo{pages}{10831} (\bibinfo{year}{2009}).

\bibitem[{\citenamefont{Cossaro et~al.}(2008)\citenamefont{Cossaro, Mazzarello,
  Rousseau, Casalis, Verdini, Kohlmeyer, Floreano, Scandolo, Morgante, Klein
  et~al.}}]{ref18}
\bibinfo{author}{\bibfnamefont{A.}~\bibnamefont{Cossaro}},
  \bibinfo{author}{\bibfnamefont{R.}~\bibnamefont{Mazzarello}},
  \bibinfo{author}{\bibfnamefont{R.}~\bibnamefont{Rousseau}},
  \bibinfo{author}{\bibfnamefont{L.}~\bibnamefont{Casalis}},
  \bibinfo{author}{\bibfnamefont{A.}~\bibnamefont{Verdini}},
  \bibinfo{author}{\bibfnamefont{A.}~\bibnamefont{Kohlmeyer}},
  \bibinfo{author}{\bibfnamefont{L.}~\bibnamefont{Floreano}},
  \bibinfo{author}{\bibfnamefont{S.}~\bibnamefont{Scandolo}},
  \bibinfo{author}{\bibfnamefont{A.}~\bibnamefont{Morgante}},
  \bibinfo{author}{\bibfnamefont{M.~L.} \bibnamefont{Klein}},
  \bibnamefont{et~al.}, \bibinfo{journal}{Science}
  \textbf{\bibinfo{volume}{321}}, \bibinfo{pages}{943} (\bibinfo{year}{2008}).

\bibitem[{\citenamefont{Wang et~al.}(2009)\citenamefont{Wang, Chi, Hush,
  Reimers, Zhang, and Ulstrup}}]{ulstrup}
\bibinfo{author}{\bibfnamefont{Y.}~\bibnamefont{Wang}},
  \bibinfo{author}{\bibfnamefont{Q.}~\bibnamefont{Chi}},
  \bibinfo{author}{\bibfnamefont{N.~S.} \bibnamefont{Hush}},
  \bibinfo{author}{\bibfnamefont{J.~R.} \bibnamefont{Reimers}},
  \bibinfo{author}{\bibfnamefont{J.}~\bibnamefont{Zhang}}, \bibnamefont{and}
  \bibinfo{author}{\bibfnamefont{J.}~\bibnamefont{Ulstrup}},
  \bibinfo{journal}{J. Phys. Chem. C} \textbf{\bibinfo{volume}{113}},
  \bibinfo{pages}{19601} (\bibinfo{year}{2009}).

\bibitem[{\citenamefont{Voznyy et~al.}(2009)\citenamefont{Voznyy, Dubowski,
  Yates, and Maksymovych}}]{voznyy}
\bibinfo{author}{\bibfnamefont{O.}~\bibnamefont{Voznyy}},
  \bibinfo{author}{\bibfnamefont{J.~J.} \bibnamefont{Dubowski}},
  \bibinfo{author}{\bibfnamefont{J.}~\bibnamefont{Yates},
  \bibfnamefont{J.~T.}}, \bibnamefont{and}
  \bibinfo{author}{\bibfnamefont{P.}~\bibnamefont{Maksymovych}},
  \bibinfo{journal}{J. Am. Chem. Soc.} \textbf{\bibinfo{volume}{131}},
  \bibinfo{pages}{12989} (\bibinfo{year}{2009}).

\bibitem[{\citenamefont{Jadzinsky et~al.}(2007)\citenamefont{Jadzinsky, Calero,
  Ackerson, Bushnell, and Kornberg}}]{ref19}
\bibinfo{author}{\bibfnamefont{P.~D.} \bibnamefont{Jadzinsky}},
  \bibinfo{author}{\bibfnamefont{G.}~\bibnamefont{Calero}},
  \bibinfo{author}{\bibfnamefont{C.~J.} \bibnamefont{Ackerson}},
  \bibinfo{author}{\bibfnamefont{D.~A.} \bibnamefont{Bushnell}},
  \bibnamefont{and} \bibinfo{author}{\bibfnamefont{R.~D.}
  \bibnamefont{Kornberg}}, \bibinfo{journal}{Science}
  \textbf{\bibinfo{volume}{318}}, \bibinfo{pages}{430} (\bibinfo{year}{2007}).

\bibitem[{\citenamefont{Walter et~al.}(2008)\citenamefont{Walter, Akola,
  Lopez-Acevedo, Jadzinsky, Calero, Ackerson, Whetten, Groenbeck, and
  Hakkinen}}]{ref20}
\bibinfo{author}{\bibfnamefont{M.}~\bibnamefont{Walter}},
  \bibinfo{author}{\bibfnamefont{J.}~\bibnamefont{Akola}},
  \bibinfo{author}{\bibfnamefont{O.}~\bibnamefont{Lopez-Acevedo}},
  \bibinfo{author}{\bibfnamefont{P.~D.} \bibnamefont{Jadzinsky}},
  \bibinfo{author}{\bibfnamefont{G.}~\bibnamefont{Calero}},
  \bibinfo{author}{\bibfnamefont{C.~J.} \bibnamefont{Ackerson}},
  \bibinfo{author}{\bibfnamefont{R.~L.} \bibnamefont{Whetten}},
  \bibinfo{author}{\bibfnamefont{H.}~\bibnamefont{Groenbeck}},
  \bibnamefont{and} \bibinfo{author}{\bibfnamefont{H.}~\bibnamefont{Hakkinen}},
  \bibinfo{journal}{Proc. Natl. Acad. Sci. (USA)}
  \textbf{\bibinfo{volume}{105}}, \bibinfo{pages}{9157} (\bibinfo{year}{2008}).

\bibitem[{\citenamefont{Strange et~al.}(2010)\citenamefont{Strange,
  Lopez-Acevedo, and H\"akkinen}}]{ref21}
\bibinfo{author}{\bibfnamefont{M.}~\bibnamefont{Strange}},
  \bibinfo{author}{\bibfnamefont{O.}~\bibnamefont{Lopez-Acevedo}},
  \bibnamefont{and}
  \bibinfo{author}{\bibfnamefont{H.}~\bibnamefont{H\"akkinen}},
  \bibinfo{journal}{J. Phys. Chem. Lett.} \textbf{\bibinfo{volume}{1}},
  \bibinfo{pages}{1528} (\bibinfo{year}{2010}).

\bibitem[{\citenamefont{Cheng et~al.}(2011)\citenamefont{Cheng, Skouta,
  Vazquez, Widawsky, Schneebeli, Chen, Hybertsen, Breslow, and
  Venkataraman}}]{cheng_nature_nanotech_2011}
\bibinfo{author}{\bibfnamefont{Z.~L.} \bibnamefont{Cheng}},
  \bibinfo{author}{\bibfnamefont{R.}~\bibnamefont{Skouta}},
  \bibinfo{author}{\bibfnamefont{H.}~\bibnamefont{Vazquez}},
  \bibinfo{author}{\bibfnamefont{J.~R.} \bibnamefont{Widawsky}},
  \bibinfo{author}{\bibfnamefont{S.}~\bibnamefont{Schneebeli}},
  \bibinfo{author}{\bibfnamefont{W.}~\bibnamefont{Chen}},
  \bibinfo{author}{\bibfnamefont{M.~S.} \bibnamefont{Hybertsen}},
  \bibinfo{author}{\bibfnamefont{R.}~\bibnamefont{Breslow}}, \bibnamefont{and}
  \bibinfo{author}{\bibfnamefont{L.}~\bibnamefont{Venkataraman}},
  \bibinfo{journal}{Nature Nanotechnology} \textbf{\bibinfo{volume}{6}},
  \bibinfo{pages}{353} (\bibinfo{year}{2011}).

\bibitem[{\citenamefont{Enkovaara et~al.}(2010)\citenamefont{Enkovaara,
  Rostgaard, Mortensen, Chen, Dulak, Ferrighi, Gavnholt, Glinsvad, Haikola,
  Hansen et~al.}}]{ref29}
\bibinfo{author}{\bibfnamefont{J.}~\bibnamefont{Enkovaara}},
  \bibinfo{author}{\bibfnamefont{C.}~\bibnamefont{Rostgaard}},
  \bibinfo{author}{\bibfnamefont{J.~J.} \bibnamefont{Mortensen}},
  \bibinfo{author}{\bibfnamefont{J.}~\bibnamefont{Chen}},
  \bibinfo{author}{\bibfnamefont{M.}~\bibnamefont{Dulak}},
  \bibinfo{author}{\bibfnamefont{L.}~\bibnamefont{Ferrighi}},
  \bibinfo{author}{\bibfnamefont{J.}~\bibnamefont{Gavnholt}},
  \bibinfo{author}{\bibfnamefont{C.}~\bibnamefont{Glinsvad}},
  \bibinfo{author}{\bibfnamefont{V.}~\bibnamefont{Haikola}},
  \bibinfo{author}{\bibfnamefont{H.~A.} \bibnamefont{Hansen}},
  \bibnamefont{et~al.}, \bibinfo{journal}{J. Phys.:Condens. Matter}
  \textbf{\bibinfo{volume}{22}}, \bibinfo{pages}{253202}
  (\bibinfo{year}{2010}).

\bibitem[{\citenamefont{Larsen et~al.}(2009)\citenamefont{Larsen, Vanin,
  Mortensen, Thygesen, and Jacobsen}}]{GPAW-LCAO}
\bibinfo{author}{\bibfnamefont{A.~H.} \bibnamefont{Larsen}},
  \bibinfo{author}{\bibfnamefont{M.}~\bibnamefont{Vanin}},
  \bibinfo{author}{\bibfnamefont{J.~J.} \bibnamefont{Mortensen}},
  \bibinfo{author}{\bibfnamefont{K.~S.} \bibnamefont{Thygesen}},
  \bibnamefont{and} \bibinfo{author}{\bibfnamefont{K.~W.}
  \bibnamefont{Jacobsen}}, \bibinfo{journal}{Phys. Rev. B}
  \textbf{\bibinfo{volume}{80}}, \bibinfo{pages}{195112}
  (\bibinfo{year}{2009}).

\bibitem[{\citenamefont{Perdew et~al.}(1996)\citenamefont{Perdew, Burke, and
  Ernzerhof}}]{ref25}
\bibinfo{author}{\bibfnamefont{J.}~\bibnamefont{Perdew}},
  \bibinfo{author}{\bibfnamefont{K.}~\bibnamefont{Burke}}, \bibnamefont{and}
  \bibinfo{author}{\bibfnamefont{M.}~\bibnamefont{Ernzerhof}},
  \bibinfo{journal}{Phys. Rev. Lett.} \textbf{\bibinfo{volume}{77}},
  \bibinfo{pages}{3865} (\bibinfo{year}{1996}).

\bibitem[{not(We use the equilibrium PBE lattice constant of 4.18~\AA~for Au.
  The distance between the second outermost Au(111) atomic surface layers in
  the left and right electrode was fixed at 21.59~\AA, 24.10~\AA~and
  26.63~\AA~for the $N$=2, 4 and 6 junctio)}]{note}
 (\bibinfo{year}{We use the equilibrium PBE lattice constant of 4.18~\AA~for
  Au. The distance between the second outermost Au(111) atomic surface layers
  in the left and right electrode was fixed at 21.59~\AA, 24.10~\AA~and
  26.63~\AA~for the $N$=2, 4 and 6 junction, respectively. The resulting
  relaxed N-Au bond length are 2.34~\AA, 2.35~\AA~and 2.33~\AA}).

\bibitem[{\citenamefont{Meir and Wingree}(1992)}]{ref30}
\bibinfo{author}{\bibfnamefont{Y.}~\bibnamefont{Meir}} \bibnamefont{and}
  \bibinfo{author}{\bibfnamefont{N.~S.} \bibnamefont{Wingree}},
  \bibinfo{journal}{Phys. Rev. Lett.} \textbf{\bibinfo{volume}{68}},
  \bibinfo{pages}{2512} (\bibinfo{year}{1992}).

\bibitem[{\citenamefont{Thygesen}(2006)}]{nonorthogonal}
\bibinfo{author}{\bibfnamefont{K.~S.} \bibnamefont{Thygesen}},
  \bibinfo{journal}{Phys. Rev. B} \textbf{\bibinfo{volume}{73}},
  \bibinfo{pages}{035309} (\bibinfo{year}{2006}).

\bibitem[{\citenamefont{Thygesen and Jacobsen}(2005)}]{wannier_thygesen}
\bibinfo{author}{\bibfnamefont{K.~S.} \bibnamefont{Thygesen}} \bibnamefont{and}
  \bibinfo{author}{\bibfnamefont{K.~W.} \bibnamefont{Jacobsen}},
  \bibinfo{journal}{Chem. Phys.} \textbf{\bibinfo{volume}{319}},
  \bibinfo{pages}{111} (\bibinfo{year}{2005}).

\bibitem[{\citenamefont{Thygesen}(2008)}]{thygesen_prl}
\bibinfo{author}{\bibfnamefont{K.~S.} \bibnamefont{Thygesen}},
  \bibinfo{journal}{Physical Review Letters} \textbf{\bibinfo{volume}{100}},
  \bibinfo{pages}{166804} (\bibinfo{year}{2008}).

\bibitem[{exp()}]{exp_butane_homo}
\urlprefix\url{http://webbook.nist.gov/cgi/cbook.cgi?ID=C106978&Units=SI&Mask=20#Ion-Energetics}.

\bibitem[{\citenamefont{Dell'Angela et~al.}(2010)\citenamefont{Dell'Angela,
  Kladnik, Cossaro, Verdini, Kamenetska, Tamblyn, Quek, Neaton, Cvetko,
  Morgante et~al.}}]{ref3}
\bibinfo{author}{\bibfnamefont{M.}~\bibnamefont{Dell'Angela}},
  \bibinfo{author}{\bibfnamefont{G.}~\bibnamefont{Kladnik}},
  \bibinfo{author}{\bibfnamefont{A.}~\bibnamefont{Cossaro}},
  \bibinfo{author}{\bibfnamefont{A.}~\bibnamefont{Verdini}},
  \bibinfo{author}{\bibfnamefont{M.}~\bibnamefont{Kamenetska}},
  \bibinfo{author}{\bibfnamefont{I.}~\bibnamefont{Tamblyn}},
  \bibinfo{author}{\bibfnamefont{S.~Y.} \bibnamefont{Quek}},
  \bibinfo{author}{\bibfnamefont{J.~B.} \bibnamefont{Neaton}},
  \bibinfo{author}{\bibfnamefont{D.}~\bibnamefont{Cvetko}},
  \bibinfo{author}{\bibfnamefont{A.}~\bibnamefont{Morgante}},
  \bibnamefont{et~al.}, \bibinfo{journal}{Nano Lett.}
  \textbf{\bibinfo{volume}{10}}, \bibinfo{pages}{2470} (\bibinfo{year}{2010}).

\bibitem[{\citenamefont{Rohlfing}(2010)}]{rohlfing}
\bibinfo{author}{\bibfnamefont{M.}~\bibnamefont{Rohlfing}},
  \bibinfo{journal}{Phys. Rev. B} \textbf{\bibinfo{volume}{82}},
  \bibinfo{pages}{205127} (\bibinfo{year}{2010}).

\bibitem[{\citenamefont{Chen}(1993)}]{stm_book}
\bibinfo{author}{\bibfnamefont{C.~J.} \bibnamefont{Chen}},
  \emph{\bibinfo{title}{Introduction to Scanning Microscopy}}
  (\bibinfo{publisher}{Oxford University Press}, \bibinfo{year}{1993}).

\bibitem[{\citenamefont{Wang et~al.}(2008)\citenamefont{Wang, Prodan, Car, and
  Selloni}}]{wang_prb_2008}
\bibinfo{author}{\bibfnamefont{J.-g.} \bibnamefont{Wang}},
  \bibinfo{author}{\bibfnamefont{E.}~\bibnamefont{Prodan}},
  \bibinfo{author}{\bibfnamefont{R.}~\bibnamefont{Car}}, \bibnamefont{and}
  \bibinfo{author}{\bibfnamefont{A.}~\bibnamefont{Selloni}},
  \bibinfo{journal}{Phys. Rev. B} \textbf{\bibinfo{volume}{77}},
  \bibinfo{pages}{245443} (\bibinfo{year}{2008}).

\bibitem[{\citenamefont{Natan et~al.}(2007)\citenamefont{Natan, Kronik, Haick,
  and Tung}}]{natan_2007}
\bibinfo{author}{\bibfnamefont{A.}~\bibnamefont{Natan}},
  \bibinfo{author}{\bibfnamefont{L.}~\bibnamefont{Kronik}},
  \bibinfo{author}{\bibfnamefont{H.}~\bibnamefont{Haick}}, \bibnamefont{and}
  \bibinfo{author}{\bibfnamefont{R.~T.} \bibnamefont{Tung}},
  \bibinfo{journal}{Adv. Mater.} \textbf{\bibinfo{volume}{19}},
  \bibinfo{pages}{4103} (\bibinfo{year}{2007}).

\bibitem[{\citenamefont{Feng et~al.}(2009)\citenamefont{Feng, Li, and
  Yang}}]{feng_jpcc_2009}
\bibinfo{author}{\bibfnamefont{X.~Y.} \bibnamefont{Feng}},
  \bibinfo{author}{\bibfnamefont{Z.}~\bibnamefont{Li}}, \bibnamefont{and}
  \bibinfo{author}{\bibfnamefont{J.}~\bibnamefont{Yang}}, \bibinfo{journal}{J.
  Phys. Chem. C} \textbf{\bibinfo{volume}{113}}, \bibinfo{pages}{21911}
  (\bibinfo{year}{2009}).

\end{thebibliography}
\end{document}